\documentclass[11pt]{article}
\usepackage[dvipsnames,table]{xcolor}
\usepackage[a4paper, left=1in, right=1in, top=1.2in, bottom=1in]{geometry}
\usepackage{setspace}
\usepackage{authblk}
\usepackage[font=footnotesize]{caption}

\usepackage{listings}
\usepackage{tikz}
\usetikzlibrary{shapes, arrows, shadows, backgrounds}
\usepackage{quantikz}
\usepackage{todonotes}
\usepackage{booktabs}
\usepackage{tabularx}
\usepackage{amsmath}
\usepackage[T1]{fontenc}

\definecolor{lightgray}{rgb}{0.9,0.9,0.9}
\definecolor{mediumgray}{rgb}{0.8,0.8,0.8}

\everymath=\expandafter{\the\everymath\displaystyle}

\lstdefinestyle{mystyle}{
	backgroundcolor=\color{white},
	commentstyle=\color{Green},
	keywordstyle=\color{blue},
	stringstyle=\color{magenta},
	numberstyle=\scriptsize\color{gray},
	basicstyle=\footnotesize,
	xleftmargin=2em,
	breaklines = true,
	captionpos = b,
	keepspaces = true,
	numbers = left,
	numbersep = 7pt,
	showspaces = false,
	showstringspaces = false,
	showtabs = false,
	tabsize = 2,
	frame=single,
	framexleftmargin=1.5em,
	xleftmargin=2em,
}
\begin{document}
\title{Extending the OmpSs-2 Programming Model for Hybrid Quantum-Classical Programming}

\author[1]{Philip Döbler}
\author[2]{David Álvarez}
\author[1]{Lucas J. Menger}
\author[1,3]{Thomas Lippert}
\author[2]{Vicenç Beltran}
\author[1]{Manpreet Singh Jattana}

\affil[1]{Modular Supercomputing and Quantum Computing, Goethe University Frankfurt, Kettenhofweg 139, 60325 Frankfurt am Main, Germany}
\affil[2]{Computer Sciences - System Tools and Advanced Runtimes, Barcelona Supercomputing Center, Plaça Eusebi Güell 1-3, 08034 Barcelona, Spain}
\affil[3]{Jülich Supercomputing Centre, Institute for Advanced Simulation, Forschungszentrum Jülich, Wilhelm-Johnen-Straße, 52428 Jülich, Germany}
\affil[ ]{\texttt{\{doebler, menger, t.lippert, jattana\}@em.uni-frankfurt.de, \{david.alvarez, vbeltran\}@bsc.es}
}

\date{February 28, 2025}

\maketitle

\begin{abstract}
    \noindent
    The OmpSs-2 programming model is used in HPC programs to parallelize code and offload code to accelerators. In this work, we extend the offloading capability to quantum computers. We explain the necessary changes to the Clang compiler and the Nanos6 runtime, which are both part of OmpSs-2. In addition, we develop a simulator that simulates a quantum computer in the network and receives the jobs offloaded by the runtime. Four detailed examples show how our programming model can be used to write hybrid quantum-classical software. The examples are random number generation, a parameter scan using the mean-field ansatz, a variational algorithm using this ansatz, and handwritten digit recognition using a hybrid convolutional neural network.
\end{abstract}

\section{Introduction}

Over the past 40 years, quantum computers have evolved from a theoretical idea~\cite{feynman_simulating_physics,universal_qc} to experiments in physics labs~\cite{qc_hw_1,qc_hw_2} to programmable machines that are becoming increasingly accessible to computer scientists~\cite{ibmq,braket}. At the same time, we are seeing a slowdown in Moore's Law, which has driven the rapid evolution of the High Performance Computing (HPC) sector for decades. To further increase the performance of HPC systems, more and more specialized accelerators are being used that are very good at solving specific types of problems. A well-known example is the GPU, which is well suited for machine learning workloads, among others. For quantum computers, it has been shown theoretically that they can solve certain types of tasks better than classical computers~\cite{shors_algorithm,grovers_algorithm}. However, they are poorly suited for other tasks that classical computers are very good at, such as the simple addition of two numbers~\cite{qc_addition,qc_addition_2}. Because of this specification, quantum computers fit very well into the accelerator model. Instead of being standalone machines, they will be used as part of a larger HPC system to handle parts of larger workloads that are well suited for them.

HPC users should therefore be able to extend their programs for quantum computers without having to also become experts in quantum mechanics. This is possible if we extend tools from the HPC world to quantum computers, so that users can continue to work in their familiar environment. One such tool is the OmpSs-2 programming model~\cite{ompss2}, which supports heterogeneous environments with accelerators such as GPUs and FPGAs. OmpSs-2 allows existing code to be parallelized and offloaded to accelerators through a set of compiler directives similar to the widely used OpenMP~\cite{openmp}. Several of the features developed in OmpSs-2 were later adopted in the OpenMP standard. The ease of use of pragma annotations, together with the already existing support for accelerators, is ideal for enabling HPC users to extend their code for quantum computing, but support for quantum accelerators is not yet available in OmpSs-2.

In this paper, we present an extension to OmpSs-2 that allows parts of the code to be offloaded to quantum computers in the form of specially declared tasks. The classical code can be written as usual in C, C++ or FORTRAN, which are widely used in HPC. Using pragma statements, we can declare kernels to be offloaded that are implemented in separate files in a domain specific quantum language. Other features of OmpSs-2, such as the task-based programming model or support for other accelerators, can be used in parallel with our quantum extension.

The remainder of the paper is organized as follows: Section~\ref{sec:related_work} gives an overview of other frameworks for quantum-classical hybrid programs. Section~\ref{sec:ompss} introduces the OmpSs-2 programming model, and in Section~\ref{sec:ompss_qpu_extension} we discuss the quantum computing extension. Section~\ref{sec:application_examples} contains four application examples that illustrate the use of our extension. The main results are concluded in Section~\ref{sec:conclusion}.

\section{Related Work}
\label{sec:related_work}
There are many programming languages specifically for quantum computing, as well as many extensions of classical programming languages to support quantum computing~\cite{qc_lang_overview_2020, qc_lang_overview_2023}. The first language was introduced in 1996~\cite{qc_lang_knill_1996} as a pseudocode to describe quantum algorithms. Review articles on this topic have already been published in the years 2004~\cite{qc_lang_overview_2004} and 2006~\cite{qc_lang_overview_2006}.

\subsection{XACC and QCOR}
The field for hybrid quantum-classical programming models however is relatively new. The first quantum-classical hybrid programming model, called eXtreme-scale ACCelerator (XACC)~\cite{xacc_1}, was presented in 2018 and described in more detail in a later publication~\cite{xacc_2}. Another paper applies the programming model to quantum chemistry~\cite{xacc_3}. XACC defines a model in which there is a host (the CPU), an accelerator (the QPU), and a buffer. The buffer is generated by the host and used by the accelerator to store results. Users describe the quantum code in kernels, which are annotated with \texttt{\_\_qpu\_\_}. The description is done at the circuit level, either in XACC assembly language (XASM) or another common language. XACC also provides a compiler to compile the quantum kernels.

The XACC framework is related to the QCOR specification~\cite{qcor_specification}. It defines how languages can be extended for single-source quantum-classical programming. The memory model defines memory spaces on the host and device. Host device memory can be allocated explicitly. The compiler allocates device memory if, for example, host memory is reserved for quantum results and handles memory transfers between host and device memory. There is a C++ implementation of the QCOR specification~\cite{qcor_compiler}. It introduces a compiler called qcor (note the small letters in contrast to the capital letters of the specification name), which provides a plugin for the clang compiler~\cite{clang} and uses the XACC framework to compile quantum resources. The compiler handles quantum-specific issues and can be invoked with similar options as classical compilers. There is also a multi threading extension to qcor~\cite{qcor_parallel}, as well as a Python implementation~\cite{qcor_python}.

\subsection{CUDA Quantum}
NVIDIA CUDA Quantum~\cite{cuda_quantum} takes a similar approach to QCOR. The extension to CUDA~\cite{cuda} also provides a single source programming model. Code for quantum computers can be written in special kernels, similar to GPU code in CUDA. There are frontends to write code in C++ and Python. To parallelize the classical parts of the code, CUDA Quantum can be combined with conventional programming models such as OpenMP~\cite{openmp}. The quantum kernels are compiled into a quantum intermediate representation that can be executed on different backends. For example, compiler flags can be used to switch between execution on a simulator and on real quantum hardware.

\subsection{QPU Extension for OpenCL}
The Open Computing Language (OpenCL)~\cite{opencl} is an open source API for heterogeneous systems. It was originally developed to program GPUs, but in the meantime offers support for additional accelerators. To execute code on an accelerator, kernels can be written in a subset of the C99 standard~\cite{c99}. These kernels can be offloaded to the accelerator via a standardized API.

Reference~\cite{opencl_qpu} extends OpenCL for QPUs. Quantum circuits are described on gate level with a syntax that resembles the C implementation of Qulacs~\cite{qulacs}. The kernel is loaded in the main program, compiled and offloaded to the QPU with a series of commands. Currently the implementation only supports simulators and no real quantum hardware. Example implementations for the Variational Quantum Eigensolver~\cite{vqe} and Shor's algorithm~\cite{shors_algorithm} can be found at~\cite{opencl_qpu_examples}.

\subsection{Quingo}
Reference~\cite{quingo} assumes a refined Heterogeneous Quantum-Classical Computation (HQCC) model for near term quantum hardware. Int this model, the hardware consists of a classical computer connected to a quantum coprocessor. The latency of the connection is so high that no real-time feedback with interleaved classical operations can take place due to the limited coherence time of the qubits. The coprocessor in turn consists of a quantum control processor that controls the qubits but can also perform classical operations, and because it is connected to the qubits with very low latency, it can be used for feedback loops in the quantum circuit and for interleaved quantum classical operations. However, the separate classical computer usually has more computing power than the quantum control processor.

The user is asked to write two programs: one program in a classical programming language such as Python or C++ that runs on the classical computer. The second program describes the quantum operations, as well as classical operations such as feedback loops that are closely related to the quantum operations and should therefore run on the quantum control processor. For the later, \cite{quingo} introduces a custom programming language that is influenced by OpenCL~\cite{opencl} called Quingo. It allows the usage of quantum as well as classical operations that are both executed on the quantum control processor. The user is responsible for minimizing classical operations on the quantum control processor and performing classical operations that are too computational extensive for the quantum control processor, on the separate classical computer. Qingo provides a compiler that compiles the code to eQASM~\cite{eqasm}, a quantum assembly language.

\subsection{QCCP}
QCCP is a task flow programming model for classical-quantum hybrid computing~\cite{qccp}. It allows functions in Python to be annotated with decorators for classical or quantum tasks. From the dependencies, QCCP builds a Direct Acyclic Graph (DAG). At runtime, there are two levels of schedulers: A workflow scheduler sends tasks to the classical and quantum backends, where they are executed by the respective local schedulers that manage the tasks on each backend. This approach is similar to our work in the sense that we also model dependencies at the task level. However, instead of creating a novel framework, we extend the well-established OmpSs-2 framework.
\section{OmpSs-2 Programming Model}
\label{sec:ompss}

OmpSs-2 is a data-flow programming model, which can be used to parallelize C, C++ and FORTRAN applications through
the use of a set of compiler directives~\cite{ompss2}.
OmpSs-2 is task-based, and a task is the minimum unit of work that can be scheduled by the programming model.
Tasks are primarily synchronized using data dependencies, which let users describe the program as a DAG of tasks.
This information is later used by the runtime, which will ensure a task execution order which guarantees the absence of
data races and the equivalence with a sequential version.
The data-flow model of OmpSs-2 is similar to OpenMP Tasking, but is oriented towards the inclusion of features that prevent the need for global synchronization points.

\begin{figure}
	\centering
	\includegraphics[width=.35\columnwidth]{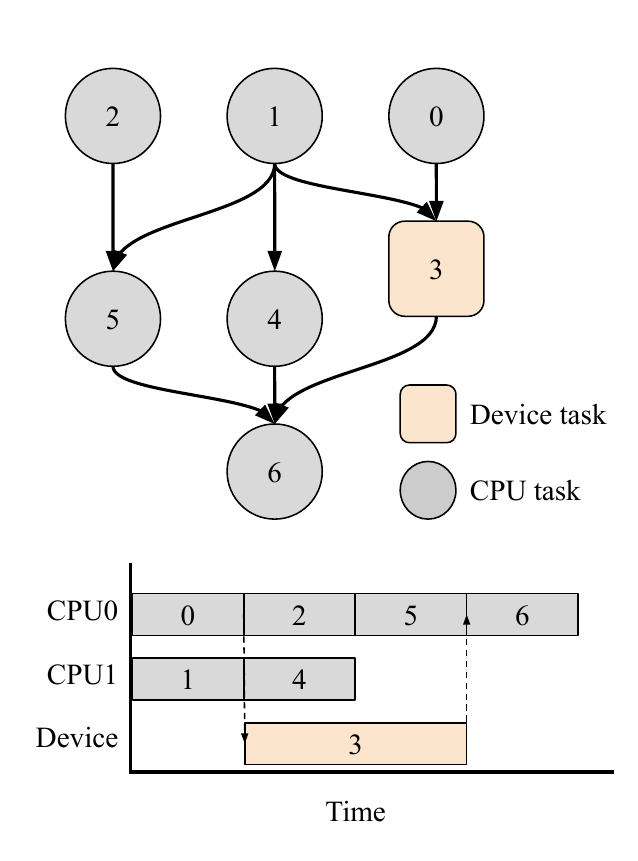}
	\caption{An example of the OmpSs-2 task graph executed on a heterogeneous platform. Tasks executing on devices (in this example, task \textit{3})
	can stablish fine-grained data dependencies with host tasks. At runtime, their offloading is overlapped with host execution, and dependencies
	cause asynchronous release of dependant tasks, as depicted by the dashed lines in the execution trace.}
	\label{fig:ompss-dev}
\end{figure}

During program execution, tasks are created online, and are submitted for execution as soon as they become ready (all of their dependencies have been satisfied).
OmpSs-2 also supports \textit{device tasks}, that are offloaded to accelerators and thus do not occupy any CPU core.
The current implementation includes support for GPUs through CUDA, and FPGAs.
Device tasks get offloaded to the appropiate accelerator when their dependencies are fulfilled, and then the OmpSs-2 runtime
will periodically check for their completion. As soon as the device task is completed, its dependencies will be released,
and any tasks which depend on their completion will be submitted for execution. This process is shown in Figure~\ref{fig:ompss-dev}, where
CPU and device tasks overlap their execution using dependencies.

Device tasks are defined in the source code by using the \texttt{device} clause in a task declaration, such as shown in Listing~\ref{lst:ompss-dev-lst}.
In this example, the \texttt{saxpy} task is defined in CUDA in a separate file, and only the declaration is included in the OmpSs-2 code, along with
a \texttt{device(cuda)} clause. The data dependencies on the buffers will synchronize the host-side \texttt{init\_buffer} tasks with the
GPU-side \texttt{saxpy} task, without requiring any explicit CUDA calls. The device task will be launched directly to an available GPU
by the OmpSs-2 runtime, which will also poll the device to determine when the task has finished.

\begin{lstlisting}[label={lst:ompss-dev-lst}, language=C, caption=\parbox{0.8\columnwidth}{OmpSs-2 device task offloading example. The saxpy function is declared as CUDA device task (line 2). When called in line 19, it is offloaded to the GPU.},float]
// Defined in kernel.cu
#pragma oss task device(cuda) in(x[n], y[n]) inout(a[n]) ndrange(...)
__global__ void saxpy(double *a, double *x, double *y, int n);

// CPU Task
#pragma oss task out(x[n])
void init_buffer(double *x, int n);

int main() {
  double *a, *x, *y;
  int n;
  // [...]
  init_buffer(a, n);
  init_buffer(x, n);
  init_buffer(y, n);

  // Saxpy will be transparently offloaded to a GPU and synchronized with the
  // init_buffer instances.
  saxpy(a, x, y, n);
#pragma oss taskwait
}
\end{lstlisting}

This data-flow mechanism for offloading tasks to accelerator allows for seamless integration in heterogeneous applications,
and offers overlapping capabilities between offloaded operations and host tasks.
This approach is the one that will be exploited in order to implement QPU offloading in OmpSs-2 programs.
\section{OmpSs-2 QPU Extension}
\label{sec:ompss_qpu_extension}

For the OmpSs-2 QPU extension, we extend the OmpSs-2 Clang compiler and the Nanos6 runtime, which together implement the OmpSs-2 programming model. The goal is to support an offloading mechanism for QPUs, similar to the already supported GPU offloading. Since computing on a quantum computer is fundamentally different from computing on a CPU or GPU, we are not trying to translate C++ code for the quantum computer. Instead, we use a widely used language for quantum computers called Open Quantum Assembly Language (OpenQASM)~\cite{openqasm}. In the C++ code, a function for the quantum computer part is declared, but not implemented, and marked with a pragma as a QPU task. The function is implemented in an OpenQASM file with the function name as the filename. This allows both traditional HPC users and quantum computing specialists to implement their respective code in a familiar language, and the two parts of the code can be easily brought together.

The OmpSs-2 interface to declare a QPU task is defined as follows:
\\\\
\setlength{\fboxsep}{5pt}
\setlength{\fboxrule}{1pt}
\fcolorbox{mediumgray}{lightgray}{%
    \parbox{.9\columnwidth}{%
        \footnotesize{\texttt{\textbf{\#pragma oss task device(qpu)}} \texttt{[clause [...]] new-line}}
		\\\footnotesize{\hphantom{8mm}\texttt{void }\texttt{\textbf{qpu\_kernel\_name}}\texttt{(int shots, qpu\_result\_t *results [,qpu\_parameters\_t *p [,const char *extension]]);}}
        }
    }
\\

The pragma to offload tasks to the QPU is \texttt{\#pragma oss task device(qpu)}. This is usually followed by declarations to model the data dependencies as shown in Section~\ref{sec:ompss} for the CUDA tasks. A function declared as a QPU task must receive at least two parameters, with two optional additional parameters, as defined in Table~\ref{tbl:params}. Moreover, the function name must match with an OpenQASM file located in a specific folder alongside the OmpSs-2 program.

\begin{table}[ht]
\centering
\caption{Parameters for QPU device offloading tasks in OmpSs-2.}
\begin{tabular}{l l l l p{180pt}}
\hline
 & Type & Direction & Mandatory & Use \\\hline\hline
0 & \texttt{int} & Input & Yes & Number of measurements the quantum computer should perform\\\hline
1 & \texttt{qpu\_results\_t *} & Output & Yes & Pointer to the results structure, containing the output from the quantum computer\\\hline
2 & \texttt{qpu\_parameters\_t *} & Input & No & Values to substitute specified parameters in the OpenQASM file\\\hline
3 & \texttt{const char *} & Input & No & String to be appended to the OpenQASM file\\\hline
\end{tabular}
\label{tbl:params}
\end{table}

The first parameter is the number of measurements the quantum computer should perform, the second parameter is a pointer to the results structure. This structure is filled by the Nanos6 runtime with the results of the quantum computer, which can then be used in the subsequent program.
In addition to the two mandatory parameters, there are two optional parameters. The first optional parameter is a structure that can contain values for parameters in the OpenQASM file. We use the OpenQASM 2.0 standard, which does not support the use of parameters by default. However, we allow the use of parameters in the OpenQASM code using the syntax \texttt{\$[x]}, where $x$ enumerates the parameters, as shown in~\cite{gaberle_thesis}. These parameters are replaced at runtime, before offloading to the quantum computer, by the values in the parameter structure passed to the function. The use of parameters whose value can be set at runtime is important, for example, for variational algorithms, as shown in the example in Section~\ref{sec:variational_algorithm}. The second optional parameter is a string that is appended to the OpenQASM code from the file. This parameter can be used either if several circuits are to be executed which differ only slightly, e.g. because measurements are made in a different basis, or if the circuit is to be changed at runtime. If there is no file with OpenQASM code or the file is empty, only this string is used. An example for the use of this extension mechanism can also be found in Section~\ref{sec:variational_algorithm}.

There is no additional parameter to set the initial state, since we consider the state preparation to be part of the quantum circuit. The initial state is either fixed and can be set in the OpenQASM code by fixed operations at the beginning of the circuit, or we have an initial state that varies at runtime, which can be achieved by parameters in the OpenQASM code. For example, we can apply the gate $R_x(\theta_i)$ to the $i$\textsuperscript{th} qubit and set either $\theta_i=0$ or $\theta_i=\pi$ at runtime, depending on whether we want to prepare the initial state \ket{0} or \ket{1} for qubit $i$.

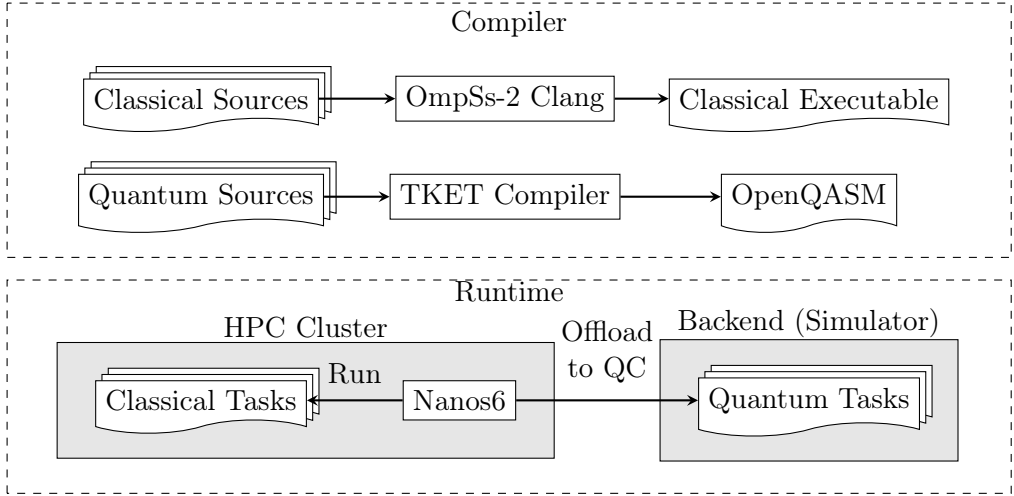
\begin{figure}[h]
	\centering
	\tikzstyle{arrow} = [thick,->,>=stealth]

\begin{tikzpicture}
	\node (classicalsources) at (0, 0) [tape, draw, tape bend top=none, double copy shadow, fill=white] {Classical Sources};
	\node (classicalcompiler) [rectangle, draw, right of=classicalsources, xshift=3cm] {OmpSs-2 Clang};
	\node (classicalexec) [tape, draw, tape bend top=none, right of=classicalcompiler, xshift=3cm] {Classical Executable};
	
	\node (quantumsources) [tape, draw, tape bend top=none, double copy shadow, fill=white, below of=classicalsources, yshift=-0.3cm] {Quantum Sources};
	\node (quantumcompiler) [rectangle, draw, right of=quantumsources, xshift=3cm] {TKET Compiler};
	\node (quantumexec) [tape, draw, tape bend top=none, right of=quantumcompiler, xshift=3cm] {OpenQASM};
	
	\draw [arrow] (classicalsources) -- (classicalcompiler);
	\draw [arrow] (classicalcompiler) -- (classicalexec);
	\draw [arrow] (quantumsources) -- (quantumcompiler);
	\draw [arrow] (quantumcompiler) -- (quantumexec);
	
	\draw[dashed] 
	($(classicalsources.north west)+(-1, 1)$) 
	rectangle 
	($(quantumexec.south east)+(1.5, -0.5)$) 
	node[midway, above, yshift=1.1cm] {Compiler};
	
	\node (classicaltasks) at (0, -4) [tape, draw, tape bend top=none, double copy shadow, fill=white] {Classical Tasks};
	\node (scheduler) [rectangle, draw, fill=white, right of=classicaltasks, xshift=2.4cm] {Nanos6};
	\node (quantumtasks) [tape, draw, tape bend top=none, double copy shadow, fill=white, right of=scheduler, xshift=3.6cm] {Quantum Tasks};
	
	\node [above=0.1cm] at ($(classicaltasks.east)!0.5!(scheduler.west)$) {Run};
	\node [above=0.1cm, align=center] at ($(scheduler.east)!0.5!(quantumtasks.west)$) {Offload\\to QC};
	
	\begin{pgfonlayer}{background}
		\draw[fill=gray!20] ($(classicaltasks.north west)+(-0.5,0.5)$) rectangle ($(scheduler.south east)+(0.5,-0.5)$) node[midway, above, yshift=0.7cm] {HPC Cluster};
		\draw[fill=gray!20] ($(quantumtasks.north west)+(-0.5,0.5)$) rectangle ($(quantumtasks.south east)+(0.5,-0.5)$) node[midway, above, yshift=0.7cm] {Backend (Simulator)};
	\end{pgfonlayer}
	
	\draw[arrow] (scheduler) -- (classicaltasks);
	\draw[arrow] (scheduler) -- (quantumtasks);
	
	\draw[dashed] 
	($(classicalsources.north west)+(-1, -5.5)$) 
	rectangle 
	($(quantumexec.south east)+(1.5, -0.8)$) 
	node[midway, above, yshift=1cm] {Runtime};
	
\end{tikzpicture}
	\caption{(Top) Compilation toolchains that process classical and quantum resources separately. The link is created by the fact that the function name in the classic code must be identical to the filename of the OpenQASM file. (Bottom) The Nanos6 runtime runs classical tasks on the HPC cluster, if their dependencies are satisfied. Quantum tasks are offloaded to the backend, which is in our case a simulator.}
	\label{fig:compilation}
\end{figure}

Figure~\ref{fig:compilation} provides a brief overview of our architecture. At the top, we see that the classical and quantum sources are processed separately by a classical and quantum compiler toolchain. The runtime is shown at the bottom. The Nanos6 scheduler executes classical tasks on the HPC cluster and offloads quantum tasks to a backend capable of processing them. In our experimental setup, the backend is a simulator, but it could be a real quantum computer.  The following sections explain the different components in more detail.

\subsection{Compiler}
We extend the OmpSs-2 clang compiler to process the pragma \texttt{device (qpu)}. For functions provided with this pragma, no implementation is expected during linking. Instead, a call to the Nanos6 runtime is inserted at the appropriate place, which takes care of offloading to the QPU. Only the classical sources are processed by the clang compiler. The quantum sources require a separate compiler toolchain. For this project, we use the TKET compiler developed by Quantinuum~\cite{tket} to compile from an OpenQASM file with arbitrary gates to an OpenQASM file with gates in the native gate set of a device. We chose the TKET compiler because of its modular structure and easy extensibility. There are several frontends to read files in different common quantum computing formats and backends can be added for custom quantum hardware or simulators. However, any other quantum compiler capable of outputting OpenQASM code can also be used.

\subsection{Runtime}
For the runtime, we extend the Nanos6 runtime of OmpSs-2 with the ability to offload OpenQASM code to a quantum computer. The Nanos6 runtime receives the name of the called function and loads the file \texttt{./qasm/<filename>.qasm}. If this file is found, its contents are loaded. If there are parameters in the OpenQASM file, they are overwritten with the given values, and if the function call has a qasm extension, it is extended to the end of the OpenQASM code.  The runtime then establishes a socket communication with the backend to send the number of shots and the circuit to be executed. The backend receives the job and returns a job ID. The runtime polls the backend for the status of all started jobs and requests the results when a job is finished.

The network connectivity of the quantum computer fits well into the loosely integrated client-server model, where the quantum hardware is a standalone machine (server) that can be accessed by multiple computing nodes (clients)~\cite{hpc_with_qpus}. Since quantum computers are currently a scarce resource and node-level integration is difficult, this model fits well with currently available hardware. If at some point a tighter integration of quantum hardware becomes possible, for example via a PCIe expansion card, the implementation in the Nanos6 runtime could be changed to accommodate the new architecture.

\subsection{Backend}
In our experimental setup, we use a simulator as the backend. The simulator acts as a quantum computer that sits on the network and can be queried at a specific IP address and port. Once connected, the simulator can be requested to do three things: receive a new job, send the status of an existing job, and return the results of a completed job. When a new job is sent, the simulator assigns a job ID and stores the OpenQASM code and the required number of shots in a dictionary. The ID is stored in a queue and also sent back to the client. A separate thread processes the jobs in the queue and stores the results. The client can use the ID to ask for the job status, which is either ``queued'', ``running'', ``completed'' or ``failed'', and also to ask for the results when the job is completed. To simulate the OpenQASM code, we use the QASM simulator from the IBM Qiskit framework~\cite{qiskit}. In this setup, a quantum compiler would not be needed, since the simulator is able to handle all commonly used gates. However, we designed the architecture with a real quantum computer in mind, and therefore, also present a solution for quantum compilation as a standard framework.

In our setup, the queue is processed according to a simple FIFO principle. This principle works well when the execution time of a quantum job is much smaller than the granularity of the batch scheduler~\cite{demistify_hpc_qc}. This is the case, for example, with superconducting quantum hardware, where gate times are on the order of tens to hundreds of nanoseconds~\cite{gate_times, ibm_devices}. For slower quantum hardware, such as trapped ion processors with gate times on the order of microseconds~\cite{gate_times}, more sophisticated scheduling methods may be useful to prevent the quantum hardware from blocking for too long. This use case is not our focus, but does not preclude use with OmpSs-2 if the backend is adapted accordingly.

\subsection{Interconnect}
The quantum computer is simulated on the same device that runs the classical program in our setup. However, we discuss what kind of interconnect would be needed for a real setup. The amount of data sent to the quantum computer depends on the size of the circuit. Today's quantum hardware can only perform a limited number of operations before the errors accumulate too much or the coherence of the qubits is lost, and the number of qubits is also limited to a few hundred at most. Hence, the circuits are currently relatively short. In our sample applications in Section~\ref{sec:application_examples}, between 10~B and 10~kB of data are sent to the quantum computer per offloading operation. The amount of data sent back depends on the number of shots and the measured output distribution. Because the backend sends the data back to the runtime in the form of a dictionary, with the measured state as the key and the number of shots as the value, the size increases as the number of different results increases. If we assume that a maximum of $10^6$ shots are used, and it is unlikely in a reasonable quantum algorithm that they are evenly distributed over all possible states, the data to be sent back remains in the range of kilobytes, maximum megabytes. The number of offloading operations can range from one to thousands. Overall, however, this does not result in data volumes that challenge the bandwidths of today's interconnects, which are in the range of hundreds of gigabits per second~\cite{prace_state_of_the_art}.

We also do not expect latency to be an issue. For example, superconducting quantum computers require tens to hundreds of nanoseconds to perform a 2-qubit operation.~\cite{gate_times,ibm_devices}. Assuming a circuit with a 2-qubit operations depth of 10, a shot takes hundreds nanoseconds to microseconds, not including the time required for setup, 1-qubit gates, and measurements.  When we run $10^3$ to $10^6$ shots, we get a run time of hundreds of microseconds to seconds, which is orders of magnitude higher than the single-digit microsecond latency of today's interconnects~\cite{prace_state_of_the_art}. Also, in many applications, we can offload multiple kernels in parallel to hide the latency. An example of this is shown in Section~\ref{sec:example_parameter_scan}.

In future applications, we can expect the required bandwidth to increase as wider and deeper quantum circuits can be run on quantum computers. For example, the famous Shor's algorithm~\cite{shors_algorithm} is expected to require something on the order of $10^{12}$ gates to break a 2048~bit RSA encryption~\cite{shor_resource_estimation}. If we assume that the description of each gate requires a few bytes in the OpenQASM file, this would mean a data transfer in the terabyte range. However, circuits of this size could be compressed very efficiently because they mostly consist of a few recurring gates. For future error correction schemes, low latency is required because the classical computations for error correction must be much faster than the decoherence times of the qubits. But we assume that error correction will be performed very close to the quantum hardware, on specialized hardware such as FPGAs, and not as part of the application implemented with OmpSs-2. We conclude that there are no exceptional requirements for the interconnection between classical and quantum hardware and that state-of-the-art solutions are well suited for this task.

\section{Application Examples}
\label{sec:application_examples}
In this section, we present four examples to show how the OmpSs-2 QPU extension can be used and the types of applications that can be realized. The first example is a minimal working example that shows the basic use of the \texttt{device(qpu)} pragma. The second example shows the use of the optional parameters explained in Section~\ref{sec:ompss_qpu_extension}. Examples three and four are more extensive and show how easily existing third-party libraries can be integrated into the program.

\subsection{Random Number Generation}
The simplest example and a kind of ``hello world'' example for quantum circuits is a coin flip. We can see the OpenQASM part of the code in Listing~\ref{lst:coin_flip_qasm}. By simply applying a Hadamard gate, we create a superposition and perform a measurement. We expect to measure the states 0 and 1 each with the probability 0.5. In the C++ code, shown in Listing~\ref{lst:coin_flip_cpp}, we define a task to print the results (line 1) and one to run the quantum code on the quantum computer (line 6), which is annotated with \texttt{device(qpu)}.

To model the dependencies between the tasks, we use the pragmas provided by OmpSs-2. For the coin flip task, the results object is an output dependency, as the task writes the results to memory. For the print results task, the results are an input dependency, as they are not modified by this task. It is the user's responsibility to fully model the dependencies with pragma annotations. We run the example with 100,000 shots and the result is 50,083 times state 0 and 49,917 times state 1. Due to the probabilistic nature of the quantum computer, this result may change from run to run.

\begin{lstlisting}[label={lst:coin_flip_qasm}, language=verilog, caption=\parbox{0.8\columnwidth}{Coin flip example OpenQASM code. The quantum computer is used to create random binary numbers.},float]
qreg q[1];
creg c[1];
h q[0];
measure q[0] -> c[0];
\end{lstlisting}

\begin{lstlisting}[label={lst:coin_flip_cpp}, language=C++, caption=\parbox{0.8\columnwidth}{Coin flip example C++ code. We annotate the function \texttt{coin\_flip} with the pragma \texttt{device(qpu)} to offload it to the quantum computer.},float]
#pragma oss task in(*results)
void print_results(qpu_result_t *results) {
  // print the results
}

#pragma oss task device(qpu) out(*results)
void coin_flip(int shots, qpu_result_t* results);

int main() {
  qpu_result_t *r = (qpu_result_t *)malloc(sizeof(qpu_result_t));
  coin_flip(NUM_SHOTS, r);
  print_results(r);
#pragma oss taskwait
}
\end{lstlisting}

\subsection{Energy Landscape Plotting}
\label{sec:example_parameter_scan}

\begin{figure}
	\centering
	\input{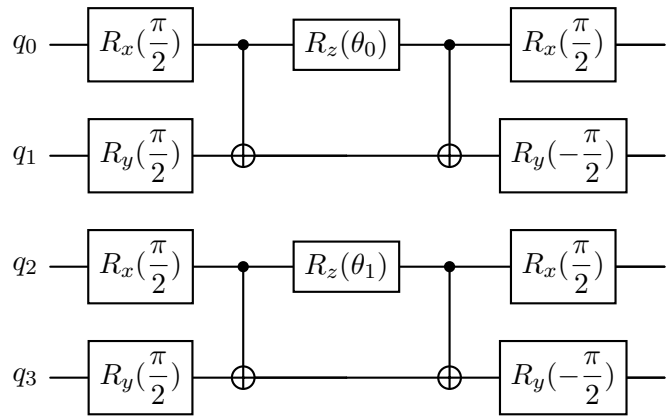}
	\caption{Circuit diagram of the mean-field ansatz for four qubits. The two parameters $\theta_0$ and $\theta_1$ can be freely selected.}
	\label{fig:ansatz_circuit_diagram}
\end{figure}

For this example we use the possibility to parameterize quantum circuits. The OpenQASM file contains a circuit called mean-field ansatz~\cite{mean-field-ansatz} with the two parameters $\theta_0$ and $\theta_1$. The circuit diagram for this example is shown in Figure~\ref{fig:ansatz_circuit_diagram}. In the OpenQASM file we use the symbol \texttt{\$[x]} for the parameters, where $x$ enumerates the parameters. Listing~\ref{lst:ansatz} shows the excerpt from the OpenQASM file of the example. We apply an $R_z$ gate to the qubits $q_0$ and $q_2$, where the rotation angles are \texttt{\$[0]} and \texttt{\$[1]}.

\begin{lstlisting}[label={lst:ansatz}, language=verilog, caption=\parbox{0.8\columnwidth}{Part of the mean-field ansatz with two parameters.},float=ht]
// ...
rz($[0]) q[0];
rz($[1]) q[2];
// ...
\end{lstlisting}

\begin{figure}[htbp]
	\centering
	\input{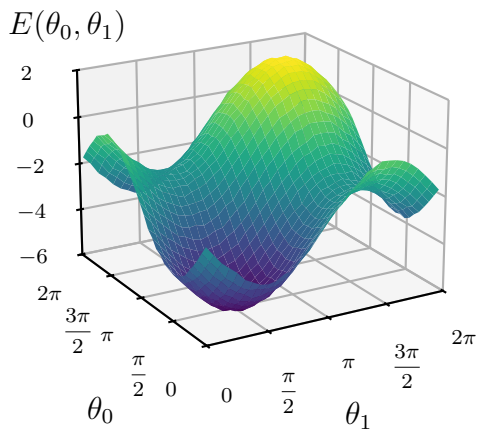}
	\caption{Expectation value $E$ for different angles $\theta_0$ and $\theta_1$.}
	\label{fig:energy_landscape}
\end{figure}

The goal of this program is to calculate the expectation value of the mean-field ansatz for different values of $\theta_0$ and $\theta_1$. To calculate the expectation value, we have to measure in different bases. This means that the same circuit is run several times with different additional rotations before the measurement. To avoid having to write several OpenQASM files that differ only slightly, we use the qasm extension feature. We iterate through an array of extensions and call the quantum computer with a different extension each time. In Listing~\ref{lst:param_scan_cpp}, line 14 we can see the definition of the extensions. We run this circuit for 32 different values of $\theta_0$ and $\theta_1$ to obtain a total of 1024 expectation values. They form an energy landscape, which is plotted in Figure~\ref{fig:energy_landscape}.

\begin{lstlisting}[label={lst:param_scan_cpp}, language=C++, caption=\parbox{0.8\columnwidth}{C++ code for the sequential parameter scan. The offloading is done sequentially.},float]
#pragma oss task device(qpu) in(*p) out(*results) in(*extension)
void ansatz(int shots, qpu_result_t *results, qpu_parameters_t *p, const char *extension);

# pragma oss task in(*results) out(*energy)
void calc_energy(qpu_result_t *results, double *energy) {
  // ...
}

int main() {
  // ...
  qpu_parameters_t *p = (qpu_parameters_t *)malloc(sizeof(qpu_parameters_t));
  // ...
  // array of qasm extensions
  const char *qasm_extensions[] = {
    "ry(pi/2) q[0];\nry(pi/2) q[1];\nmeasure q[0] -> c[0];\nmeasure q[1] -> c[1];",
    "ry(pi/2) q[0];\nry(pi/2) q[2];\nmeasure q[0] -> c[0];\nmeasure q[2] -> c[2];",
    "ry(pi/2) q[0];\nry(pi/2) q[3];\nmeasure q[0] -> c[0];\nmeasure q[3] -> c[3];",
    // ...
  }
  initialize_parameters(p); 
  double *energy = (double *)malloc(sizeof(double));
  double step_size = (2 * M_PI) / steps;
  for (size_t i = 0; i < steps; i++) {
    p->parameters[0] = i * step_size;
    for (size_t j = 0; j < steps; j++) {
      p->parameters[1] = j * step_size;
      double total_energy = 0.;
      for (size_t k = 0; k < num_ansaetze; k++) {
        ansatz(num_shots, r, p, qasm_extensions[k]);
        calc_energy(r, energy);
#pragma oss taskwait
        total_energy += *energy;
      }
      // print energy for current parameters
    }
  }
  // free resources
}
\end{lstlisting}

The code in Listing~\ref{lst:param_scan_cpp} offloads the circuits sequentially. It means that the circuit with the first OpenQASM extension is offloaded, and only after the results are sent back from the backend and further processed in the \texttt{calc\_energy} function, the next circuit is offloaded. Since we are simulating a single quantum computer in our backend, this approach is feasible. However, the strength of OmpSs-2 is that we can do the offloading asynchronously. This can help to hide the latency even if there is only one quantum computer processing the circuits. But the main advantage is that in a setup with multiple quantum computers, the circuits are processed in parallel.

In Listing~\ref{lst:param_scan_parallel_cpp}, we can see a modified version of the code. We create as many result struct objects as there are OpenQASM extensions (line 11). The taskwait pragma is not called inside the inner loop, instead we model the dependencies for the parameters and the total energy. The \texttt{calc\_energy} function is declared as a task that performs a reduction by summing the energy values of the 18 different circuits (line 2). This means that once the first quantum circuit is offloaded, the next one can be sent directly to the quantum computer without waiting for the results of the first one. The taskwait pragma on line 49 prevents the main function from exiting before all tasks have finished. In our experimental setup, we only simulate one quantum computer. Nevertheless the runtime for the parameter scan is reduced by 14\%, if we use the parallel version, since we can hide the latency. This improvement certainly depends on the size of the circuits, as well as the latency and bandwidths of the connection between runtime and backend.

\begin{lstlisting}[label={lst:param_scan_parallel_cpp}, language=C++, caption=\parbox{0.8\columnwidth}{C++ code for the parallel parameter scan (see results in Figure~\ref{fig:energy_landscape}). The offloading in line 26 is done in parallel for all OpenQASM extensions.},float]
// ...
#pragma oss task in(*results) reduction(+: [1]energy)
void calc_energy(qpu_result_t *results, double *energy) {
  // ...
}

int main() {
  // ...
  qpu_parameters_t *p = (qpu_parameters_t *)malloc(sizeof(qpu_parameters_t));
  // define a result struct for every ansatz
  qpu_result_t *r[num_ansaetze];
  for (size_t i = 0; i < num_ansaetze; i++) {
    r[i] = (qpu_result_t *)malloc(sizeof(qpu_result_t));
  }

  // array of qasm extensions
  // ...

  // array of energies
  double *energy[num_ansaetze];
  for (size_t i = 0; i < num_ansaetze; i++) {
    energy[i] = (double *)malloc(sizeof(double));
  }
  // ...
  for (size_t i = 0; i < steps; i++) {
#pragma oss task out(*p)
    {
      p->parameters[0] = i * step_size;
    }
    for (size_t j = 0; j < steps; j++) {
#pragma oss task out(*p)
      {
        p->parameters[1] = j * step_size;
      }      
#pragma oss task out(total_energy)
      {
        total_energy = 0.0f;
      }
      for (size_t k = 0; k < num_ansaetze; k++) {
        ansatz(num_shots, r[k], p, qasm_extensions[k]);
        calc_energy(r[k], energy[k]);
      }
#pragma oss task in(total_energy)
      {
        // print energy
      }
    }
  }
#pragma oss taskwait
}
\end{lstlisting}

\subsection{Variational Algorithm}
\label{sec:variational_algorithm}

\begin{figure}[htbp]
	\centering
	\begin{tikzpicture}[node distance=3cm,auto,every node/.style={font=\footnotesize}]
    \node[draw, ellipse, minimum width=1.4cm, minimum height=0.7cm] (cpu) {CPU};
    \node[draw, ellipse, minimum width=1.4cm, minimum height=0.7cm, right of=cpu] (qpu) {QPU};

    \draw[-Latex, thick] (qpu) to[out=150, in=30] node[midway, above, align=center] {measured\\bitstrings} (cpu);
    \draw[-Latex, thick] (cpu) to[out=-30, in=-150] node[midway, below, align=center] {updated\\parameters} (qpu);
\end{tikzpicture}
	\caption{Schematic representation of the execution of variational algorithms. The CPU optimizes the parameters of the quantum circuit, the QPU returns the bit strings of the current measurements.}
	\label{fig:variational_scheme}
\end{figure}
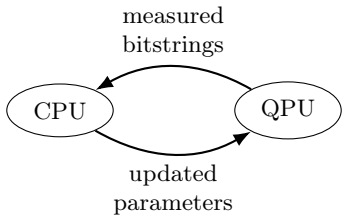

In variational algorithms, a circuit called ansatz is implemented on the quantum computer. The goal of the algorithm is to find the lowest expectation value of $\bra{\Psi(\theta)} H \ket{\Psi(\theta)}$, where $\Psi(\theta)$ is the parametrized ansatz and $H$ is the problem Hamiltonian. To do this, we use a classical optimization algorithm on classical hardware that adjusts the parameters and offloads the execution of the ansatz onto the quantum computer. In Figure~\ref{fig:variational_scheme}, we see that this results in a loop in which updated parameters are sent to the quantum computer and the measured bit strings are returned. Variational algorithms play an important role in the Noisy Intermediate Scale Quantum (NISQ) era~\cite{nisq_era} because they allow problems to be solved with shorter and therefore less error-prone circuits. They are used, for example, to find the ground state energy of molecules~\cite{vqe_original,vqe_overview} or to solve combinatorial optimization problems~\cite{qaoa_original,qaoa_review}.

In this example, we want to find the lowest expectation value of the circuit from Section~\ref{sec:example_parameter_scan}. To do this, we use the widely used NLopt library~\cite{nlopt}, which contains a large number of optimization algorithms. Listing~\ref{lst:variational_algo_cpp} shows the C++ code for this example. In the main function we set the optimization algorithm, DIviding RECTangles (DIRECT)~\cite{optimization_direct} in our case and set the objective function. We also define bounds for the parameters from 0 to $2\pi$ and set a stopping criteria (not shown). In the objective function, we call the quantum computer and calculate the expectation value similar to the example in Section~\ref{sec:example_parameter_scan}. Calling \texttt{opt.optimize} starts the optimizer which calls the objective function, which in turn calls the quantum computer. This example shows how easy it is to use 3rd party libraries with OmpSs-2. With the quantum extension, it is also easy to outsource parts of existing algorithms to a quantum computer without having to implement the algorithms from scratch.

We run the example with 10,000 shots per evaluation of the circuit. The optimizer finds a minimum of $-5.94$. Theoretically the minimum is $-6$. To find a more precise value, we can increase the number of shots and decrease the relative tolerance of the stopping criterion. However, this also increases run time.

\begin{lstlisting}[label={lst:variational_algo_cpp}, language=C++, caption=\parbox{0.8\columnwidth}{C++ part of the variational algorithm. We use the NLopt library for optimization.},float]
// ...
#pragma oss task device(qpu) in(*p) out(*results) in(*extension)
void ansatz(int shots, qpu_result_t *results, qpu_parameters_t *p, const char *extension);

#pragma oss task in(*results) out(*energy)
void calc_energy(qpu_result_t *results, double *energy) {
  // ...
}

double evaluate_energy(const std::vector<double> &x, std::vector<double> &grad, void *func_data) {
  // ...
  for (size_t k = 0; k < num_ansaetze; k++) {
    ansatz(num_shots, r, p, qasm_extensions[k]);
    calc_energy(r, energy);
#pragma oss taskwait
    total_energy += *energy;
  }
  // ...
  return total_energy;
}

int main() {
  nlopt::opt opt(nlopt::GN_DIRECT_L, 2);
  // ...
  opt.set_min_objective(evaluate_energy, NULL);
  // run optimization and print result
  try {
    nlopt::result result = opt.optimize(x, minf);
    // print result
  } catch (std::exception &e) {
    // print error
  }
}
\end{lstlisting}

\subsection{Quantum Convolution}

Quantum machine learning (QML) is a highly active area of research with ongoing advancements~\cite{Garcia2024}. Although the potential for quantum advantage in machine learning remains uncertain, it serves as an intriguing example to demonstrate our programming model. The size of modern neural networks has increased dramatically, leading to the conclusion that fully QML algorithms are unlikely to emerge in the near future~\cite{Garcia2024}. Instead, hybrid quantum-classical algorithms are more plausible. In such hybrid approaches, it is crucial to efficiently distribute the computational load across various subsystems, such as CPUs, GPUs, and, in this case, QPUs. 

To evaluate the applicability of the OmpSs-2 extension in QML, we replicate the quantum computing convolutional neural network described in~\cite{Matic2022}. This example provides a valuable testing opportunity, involving rapid and frequent interleaved computations between classical and quantum hardware. While the example itself features a small neural network that does not require HPC infrastructure, the ever-growing computational demands of neural networks in general are shifting their execution and training into the realm of HPC, as evidenced by the development of new AI-focused supercomputers like Jupiter~\cite{Dumiak2023,HertenSC24}. The OmpSs-2 framework is designed to operate in such environments, making this extension an ideal fit for large-scale and scalable QML algorithms in the future. 

\subsubsection{Network Architecture and Implementation}

\begin{figure}
	\centering
	\includegraphics[width=0.9\columnwidth]{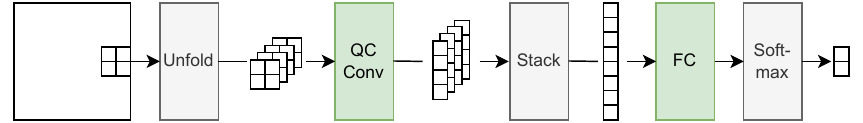}
	\caption{The image depicts a process where the input image is divided into 2x2 patches, which translates to a kernel size of (2,2) and a stride of 2. A quantum convolution (QC Conv) is applied to each patch, generating 4 features per patch. For quantum convolution we use the higher order encoding basic entangling layer outlined in~\cite{Matic2022}. These features are stacked into a single tensor, which is then processed by a fully connected layer (FC) to map the features to prediction classes, followed by a softmax activation for the output.}
	\label{fig:network_architecture}
\end{figure}

Convolutional Neural Networks (CNNs)~\cite{LeCun1995,Krizhevsky2012} are a class of deep learning models widely used for tasks such as image recognition. A CNN typically consists of convolutional layers, pooling layers, and fully connected layers. The convolutional layer relies on trainable kernel matrices to extract features from images by convolving these filters over the input data. The polling layer summarizes data, and the fully connected layer serves as the classifier, combining the learned features to produce the output. In the quantum version, the convolutional operation is translated into a quantum convolution, where a quantum circuit is designed to generate image features by leveraging the principles of quantum computing. The architecture of the network, including the quantum convolutional component, is depicted in Figure~\ref{fig:network_architecture}, providing an overview of how classical and quantum computations are integrated within the model. 

Similar to~\cite{Matic2022}, we simulate the quantum circuit. For simplicity, the training of the neural network is implemented in Python using the PyTorch~\cite{PyTorch2019} package. Unlike~\cite{Matic2022}, we implemented the circuit simulation as PyTorch module. This approach involves calculating the unitary matrix corresponding to the quantum circuit and then determining the resulting probabilistic state vector. This implementation enables highly parallelized training of the network on GPUs, significantly improving efficiency. The enhanced efficiency allows us to test the network on much larger datasets. 

\subsubsection{Training and Comparison}

For the training dataset, we use the widely known dataset MNIST CIFAR-10~\cite{Deng2012}. As a reference to~\cite{Matic2022}, we briefly summarize the performance of the quantum CNN and a classical CNN. The classical CNN has the same structure as the quantum CNN, but the convolution operation is a standard classical convolution with the same kernel size, stride, and four output channels. The other difference is that the output is activated using the rectified linear unit (ReLU) function. 

All networks are trained using the ADAM optimizer~\cite{adam} with the following hyper parameters: learning rate $\gamma = 0.001$, $\beta_1 = 0.9$, $\beta_2 = 0.999$, and  $\epsilon = 10^{-8}$. Training is conducted for 50 epochs, repeated 20 times, and the optimal number of epochs is selected based on the best average performance on the validation dataset. The final evaluation is performed on the test dataset.  For the classical CNN, we obtain an average out-of-sample accuracy of 95.77\% with a standard deviation of 2.02\%. The quantum CNN performs slightly worse with an average out-of-sample accuracy of 90.58\% and a standard deviation of 5.05\%.

\subsubsection{Implementation with OmpSs-2}

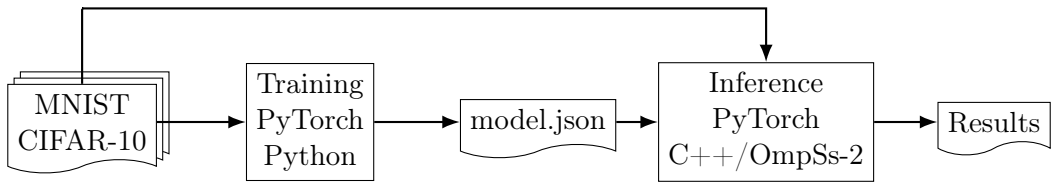
\begin{figure}[htbp]
    \centering
    \begin{tikzpicture}[node distance=3cm]
    \node(mnist)[double copy shadow, tape, draw=black, fill=white, tape bend top=none, align=center] {MNIST\\CIFAR-10};
    \node(training)[right of=mnist, rectangle, draw, align=center]{Training\\PyTorch\\Python};
    \node(model)[right of=training, tape, draw=black, fill=white, tape bend top=none, align=center]{model.json};
    \node(inference)[right of=model, rectangle, draw, align=center]{Inference\\PyTorch\\C++/OmpSs-2};
    \node(results)[right of=inference, tape, draw=black, fill=white, tape bend top=none, align=center]{Results};
    \node(dummy)[above of=model, draw=none]{};

\draw[-Latex, thick] (mnist) -- (training);
\draw[-Latex, thick] (training) -- (model);
\draw[-Latex, thick] (model) -- (inference);
\draw[-Latex, thick] (inference) -- (results);
\node (midpoint) at ($(mnist.north) + (0,1cm)$) {};
\draw[-Latex, thick] (mnist) |- ($(mnist.north) + (0,1cm)$) -| (inference);

\end{tikzpicture}
    \caption{High level overview of the QML example. The training is done with the PyTorch Python frontend. The resulting model is written to a json file, which is read into the C++ implementation that uses OmpSs-2 for offloading to the quantum computer.}
    \label{fig:qml_overview}
\end{figure}

We use PyTorch's C++ frontend to test one of the trained networks with OmpSs-2. Figure~\ref{fig:qml_overview} gives a high-level overview of our setup. In the Python code, we store the network parameters of the trained network in a json file. We read the model data of the fully trained network into the C++ implementation and do only the inference with OmpSs-2. To do this, we implement a network consisting of a standard fully connected layer and a custom quantum convolution layer. This quantum convolution layer makes a call to the quantum computer in its forward function. We evaluate this network on the test dataset and get an accuracy of 95.02\%, which is in the expected range. This example shows that it is possible to implement more extensive examples that rely heavily on third-party libraries such as PyTorch.

\section{Conclusion and Future Work}
\label{sec:conclusion}
We presented an extension of the OmpSs-2 framework to support quantum computers. The extension allows to offload functions from the classical code to a networked quantum computer. The quantum code is written in OpenQASM, which is well established in the field of quantum computing, while the HPC code can be kept in a classical HPC language such as C++. Therefore, our solution provides a modular approach to hybrid quantum-classical computing that allows both quantum and HPC specialists to work in their familiar environments. Portions of the existing classical code can be incrementally ported to the quantum computer without having to completely rewrite the code. Only the quantum part needs to be implemented in a quantum-specific language. We showed four applications of different sizes to demonstrate the strength of the framework. 

For future work, we want to improve performance through various measures. The OpenQASM file could be preloaded at the start of the program and not only when the function is actually called. It could also be reused instead of being reloaded from the file when the same function is called again. We also plan to introduce an abstraction layer for the backends to allow easier support for new quantum computers. To make the framework easier to use, we would like to provide the structures for storing results and parameters with convenience functions. A further step could be to develop a library for a specific use case, such as variational algorithms, that contains the necessary OpenQASM and C++ code. In the current NISQ era, error mitigation~\cite{em_original,em_overview} plays an important role to improve the results obtained from the error-prone device via post-processing. It would be convenient for users to implement the classical and quantum parts, for example, of matrix-based error mitigation~\cite{em_matrix_based_1,em_matrix_based_2,em_matrix_based_3} as a function for users to call. So far, the framework has only been tested on a simulator. In Section~\ref{sec:example_parameter_scan} we showed an example, where the execution of multiple kernels can be parallelized. It would be interesting to test this code in an environment with several simulators or real devices to see how well the parallelization works.  There could also be a configuration where there are interdependencies between several different types of quantum computers~\cite{triple_hybrid}. Such a setup would be interesting to test, as it would take advantage of OmpSs-2 strengths in modeling dependencies.

\bibliographystyle{IEEEtran}
\bibliography{references}

\begin{thebibliography}{10}
\providecommand{\url}[1]{#1}
\csname url@samestyle\endcsname
\providecommand{\newblock}{\relax}
\providecommand{\bibinfo}[2]{#2}
\providecommand{\BIBentrySTDinterwordspacing}{\spaceskip=0pt\relax}
\providecommand{\BIBentryALTinterwordstretchfactor}{4}
\providecommand{\BIBentryALTinterwordspacing}{\spaceskip=\fontdimen2\font plus
\BIBentryALTinterwordstretchfactor\fontdimen3\font minus
  \fontdimen4\font\relax}
\providecommand{\BIBforeignlanguage}[2]{{%
\expandafter\ifx\csname l@#1\endcsname\relax
\typeout{** WARNING: IEEEtran.bst: No hyphenation pattern has been}%
\typeout{** loaded for the language `#1'. Using the pattern for}%
\typeout{** the default language instead.}%
\else
\language=\csname l@#1\endcsname
\fi
#2}}
\providecommand{\BIBdecl}{\relax}
\BIBdecl

\bibitem{feynman_simulating_physics}
\BIBentryALTinterwordspacing
R.~P. Feynman, ``Simulating physics with computers,'' \emph{International
  Journal of Theoretical Physics}, vol.~21, no.~6, pp. 467--488, 06 1982.
  [Online]. Available: \url{https://doi.org/10.1007/BF02650179}
\BIBentrySTDinterwordspacing

\bibitem{universal_qc}
D.~Deutsch, ``Quantum theory, the church--turing principle and the universal
  quantum computer,'' \emph{Proceedings of the Royal Society of London. A.
  Mathematical and Physical Sciences}, vol. 400, no. 1818, pp. 97--117, 1985.

\bibitem{qc_hw_1}
J.~A. Jones and M.~Mosca, ``Implementation of a quantum algorithm on a nuclear
  magnetic resonance quantum computer,'' \emph{The Journal of chemical
  physics}, vol. 109, no.~5, pp. 1648--1653, 1998.

\bibitem{qc_hw_2}
\BIBentryALTinterwordspacing
I.~L. Chuang, N.~Gershenfeld, and M.~Kubinec, ``Experimental implementation of
  fast quantum searching,'' \emph{Phys. Rev. Lett.}, vol.~80, pp. 3408--3411,
  Apr 1998. [Online]. Available:
  \url{https://link.aps.org/doi/10.1103/PhysRevLett.80.3408}
\BIBentrySTDinterwordspacing

\bibitem{ibmq}
\BIBentryALTinterwordspacing
{IBM}, ``{IBM Quantum},'' 2025. [Online]. Available:
  \url{https://quantum.ibm.com/}
\BIBentrySTDinterwordspacing

\bibitem{braket}
\BIBentryALTinterwordspacing
{Amazon Web Services}, ``{Amazon Braket},'' 2025. [Online]. Available:
  \url{https://aws.amazon.com/braket/}
\BIBentrySTDinterwordspacing

\bibitem{shors_algorithm}
P.~W. Shor, ``Algorithms for quantum computation: discrete logarithms and
  factoring,'' in \emph{Proceedings 35th annual symposium on foundations of
  computer science}.\hskip 1em plus 0.5em minus 0.4em\relax Ieee, 1994, pp.
  124--134.

\bibitem{grovers_algorithm}
\BIBentryALTinterwordspacing
L.~K. Grover, ``A fast quantum mechanical algorithm for database search,'' in
  \emph{Proceedings of the Twenty-Eighth Annual ACM Symposium on Theory of
  Computing}, ser. STOC '96.\hskip 1em plus 0.5em minus 0.4em\relax New York,
  NY, USA: Association for Computing Machinery, 1996, p. 212–219. [Online].
  Available: \url{https://doi.org/10.1145/237814.237866}
\BIBentrySTDinterwordspacing

\bibitem{qc_addition}
\BIBentryALTinterwordspacing
T.~G. Draper, ``Addition on a quantum computer,'' 2000. [Online]. Available:
  \url{https://arxiv.org/abs/quant-ph/0008033}
\BIBentrySTDinterwordspacing

\bibitem{qc_addition_2}
\BIBentryALTinterwordspacing
K.~Michielsen, M.~Nocon, D.~Willsch, F.~Jin, T.~Lippert, and H.~{De Raedt},
  ``Benchmarking gate-based quantum computers,'' \emph{Computer Physics
  Communications}, vol. 220, pp. 44--55, 2017. [Online]. Available:
  \url{https://www.sciencedirect.com/science/article/pii/S0010465517301935}
\BIBentrySTDinterwordspacing

\bibitem{ompss2}
J.~M. Perez, V.~Beltran, J.~Labarta, and E.~Ayguadé, ``Improving the
  integration of task nesting and dependencies in openmp,'' in \emph{2017 IEEE
  International Parallel and Distributed Processing Symposium (IPDPS)}, 2017,
  pp. 809--818.

\bibitem{openmp}
L.~Dagum and R.~Menon, ``Openmp: an industry standard api for shared-memory
  programming,'' \emph{IEEE Computational Science and Engineering}, vol.~5,
  no.~1, pp. 46--55, 1998.

\bibitem{qc_lang_overview_2020}
\BIBentryALTinterwordspacing
B.~Heim, M.~Soeken, S.~Marshall, C.~Granade, M.~Roetteler, A.~Geller,
  M.~Troyer, and K.~Svore, ``Quantum programming languages,'' \emph{Nature
  Reviews Physics}, vol.~2, no.~12, pp. 709--722, Dec 2020. [Online].
  Available: \url{https://doi.org/10.1038/s42254-020-00245-7}
\BIBentrySTDinterwordspacing

\bibitem{qc_lang_overview_2023}
S.~Lopez~Alarcón, E.~Wong, T.~S. Humble, and E.~Dumitrescu, ``Quantum
  programming paradigms and description languages,'' \emph{Computing in Science
  \& Engineering}, vol.~25, no.~6, pp. 33--38, 2023.

\bibitem{qc_lang_knill_1996}
\BIBentryALTinterwordspacing
E.~Knill, ``Conventions for quantum pseudocode,'' 6 1996. [Online]. Available:
  \url{https://www.osti.gov/biblio/366453}
\BIBentrySTDinterwordspacing

\bibitem{qc_lang_overview_2004}
P.~Selinger, ``A brief survey of quantum programming languages,'' in
  \emph{Functional and Logic Programming}, Y.~Kameyama and P.~J. Stuckey,
  Eds.\hskip 1em plus 0.5em minus 0.4em\relax Berlin, Heidelberg: Springer
  Berlin Heidelberg, 2004, pp. 1--6.

\bibitem{qc_lang_overview_2006}
S.~J. GAY, ``Quantum programming languages: survey and bibliography,''
  \emph{Mathematical Structures in Computer Science}, vol.~16, no.~4, p.
  581–600, 2006.

\bibitem{xacc_1}
\BIBentryALTinterwordspacing
A.~McCaskey, E.~Dumitrescu, D.~Liakh, M.~Chen, W.~Feng, and T.~Humble, ``A
  language and hardware independent approach to quantum–classical
  computing,'' \emph{SoftwareX}, vol.~7, pp. 245--254, 2018. [Online].
  Available:
  \url{https://www.sciencedirect.com/science/article/pii/S2352711018300700}
\BIBentrySTDinterwordspacing

\bibitem{xacc_2}
\BIBentryALTinterwordspacing
A.~J. McCaskey, D.~I. Lyakh, E.~F. Dumitrescu, S.~S. Powers, and T.~S. Humble,
  ``Xacc: a system-level software infrastructure for heterogeneous
  quantum–classical computing*,'' \emph{Quantum Science and Technology},
  vol.~5, no.~2, p. 024002, feb 2020. [Online]. Available:
  \url{https://dx.doi.org/10.1088/2058-9565/ab6bf6}
\BIBentrySTDinterwordspacing

\bibitem{xacc_3}
\BIBentryALTinterwordspacing
D.~Claudino, A.~J. McCaskey, and D.~I. Lyakh, ``A backend-agnostic,
  quantum-classical framework for simulations of chemistry in c++,'' vol.~4,
  no.~1, oct 2022. [Online]. Available: \url{https://doi.org/10.1145/3523285}
\BIBentrySTDinterwordspacing

\bibitem{qcor_specification}
\BIBentryALTinterwordspacing
T.~M. Mintz, A.~J. McCaskey, E.~F. Dumitrescu, S.~V. Moore, S.~Powers, and
  P.~Lougovski, ``Qcor: A language extension specification for the
  heterogeneous quantum-classical model of computation,'' \emph{J. Emerg.
  Technol. Comput. Syst.}, vol.~16, no.~2, mar 2020. [Online]. Available:
  \url{https://doi.org/10.1145/3380964}
\BIBentrySTDinterwordspacing

\bibitem{qcor_compiler}
\BIBentryALTinterwordspacing
A.~Mccaskey, T.~Nguyen, A.~Santana, D.~Claudino, T.~Kharazi, and H.~Finkel,
  ``Extending c++ for heterogeneous quantum-classical computing,'' \emph{ACM
  Transactions on Quantum Computing}, vol.~2, no.~2, jul 2021. [Online].
  Available: \url{https://doi.org/10.1145/3462670}
\BIBentrySTDinterwordspacing

\bibitem{clang}
C.~team, ``Clang: a c language family frontend for llvm,'' 2024,
  \url{https://docs.rigetti.com/qcs} [Accessed: Aug 28, 2024].

\bibitem{qcor_parallel}
A.~Hayashi, A.~Adams, J.~Young, A.~McCaskey, E.~Dumitrescu, V.~Sarkar, and
  T.~M. Conte, ``Enabling multi-threading in heterogeneous quantum-classical
  programming models,'' in \emph{2023 IEEE International Parallel and
  Distributed Processing Symposium Workshops (IPDPSW)}, 2023, pp. 509--516.

\bibitem{qcor_python}
\BIBentryALTinterwordspacing
T.~Nguyen and A.~J. McCaskey, ``Extending python for quantum-classical
  computing via quantum just-in-time compilation,'' \emph{ACM Transactions on
  Quantum Computing}, vol.~3, no.~4, jul 2022. [Online]. Available:
  \url{https://doi.org/10.1145/3544496}
\BIBentrySTDinterwordspacing

\bibitem{cuda_quantum}
J.-S. Kim, A.~McCaskey, B.~Heim, M.~Modani, S.~Stanwyck, and T.~Costa, ``Cuda
  quantum: The platform for integrated quantum-classical computing,'' in
  \emph{2023 60th ACM/IEEE Design Automation Conference (DAC)}, 2023, pp. 1--4.

\bibitem{cuda}
J.~Nickolls, I.~Buck, M.~Garland, and K.~Skadron, ``Scalable parallel
  programming with cuda: Is cuda the parallel programming model that
  application developers have been waiting for?'' \emph{Queue}, vol.~6, no.~2,
  pp. 40--53, 2008.

\bibitem{opencl}
J.~E. Stone, D.~Gohara, and G.~Shi, ``Opencl: A parallel programming standard
  for heterogeneous computing systems,'' vol.~12, no.~3, p. 66–73, may 2010.

\bibitem{c99}
\BIBentryALTinterwordspacing
ISO, \emph{C99 Standard}, 1999, iSO/IEC 9899:1999. [Online]. Available:
  \url{https://www.iso.org/standard/29237.html}
\BIBentrySTDinterwordspacing

\bibitem{opencl_qpu}
J.~V{\'a}zquez-P{\'e}rez, C.~Pi{\~n}eiro, J.~C. Pichel, T.~F. Pena, and
  A.~G{\'o}mez, ``Qpu integration in opencl for heterogeneous programming,''
  \emph{The Journal of Supercomputing}, pp. 1--22, 2024.

\bibitem{qulacs}
\BIBentryALTinterwordspacing
Y.~Suzuki, Y.~Kawase, Y.~Masumura, Y.~Hiraga, M.~Nakadai, J.~Chen, K.~M.
  Nakanishi, K.~Mitarai, R.~Imai, S.~Tamiya, T.~Yamamoto, T.~Yan, T.~Kawakubo,
  Y.~O. Nakagawa, Y.~Ibe, Y.~Zhang, H.~Yamashita, H.~Yoshimura, A.~Hayashi, and
  K.~Fujii, ``Qulacs: a fast and versatile quantum circuit simulator for
  research purpose,'' \emph{{Quantum}}, vol.~5, p. 559, Oct. 2021. [Online].
  Available: \url{https://doi.org/10.22331/q-2021-10-06-559}
\BIBentrySTDinterwordspacing

\bibitem{vqe}
\BIBentryALTinterwordspacing
A.~Peruzzo, J.~McClean, P.~Shadbolt, M.-H. Yung, X.-Q. Zhou, P.~J. Love,
  A.~Aspuru-Guzik, and J.~L. O'Brien, ``A variational eigenvalue solver on a
  photonic quantum processor,'' \emph{Nature Communications}, vol.~5, no.~1, p.
  4213, Jul 2014. [Online]. Available: \url{https://doi.org/10.1038/ncomms5213}
\BIBentrySTDinterwordspacing

\bibitem{opencl_qpu_examples}
J.~V{\'a}zquez-P{\'e}rez. (2024) Openclqpu samples. Available:
  \url{https://github.com/jorgevazquezperez/OpenCLQPU/tree/main/samples}.

\bibitem{quingo}
\BIBentryALTinterwordspacing
X.~Fu, J.~Yu, X.~Su, H.~Jiang, H.~Wu, F.~Cheng, X.~Deng, J.~Zhang, L.~Jin,
  Y.~Yang, L.~Xu, C.~Hu, A.~Huang, G.~Huang, X.~Qiang, M.~Deng, P.~Xu, W.~Xu,
  W.~Liu, Y.~Zhang, Y.~Deng, J.~Wu, and Y.~Feng, ``Quingo: A programming
  framework for heterogeneous quantum-classical computing with nisq features,''
  \emph{ACM Transactions on Quantum Computing}, vol.~2, no.~4, dec 2021.
  [Online]. Available: \url{https://doi.org/10.1145/3483528}
\BIBentrySTDinterwordspacing

\bibitem{eqasm}
X.~Fu, L.~Riesebos, M.~A. Rol, J.~van Straten, J.~van Someren, N.~Khammassi,
  I.~Ashraf, R.~F.~L. Vermeulen, V.~Newsum, K.~K.~L. Loh, J.~C. de~Sterke,
  W.~J. Vlothuizen, R.~N. Schouten, C.~G. Almudever, L.~DiCarlo, and
  K.~Bertels, ``eqasm: An executable quantum instruction set architecture,'' in
  \emph{2019 IEEE International Symposium on High Performance Computer
  Architecture (HPCA)}, 2019, pp. 224--237.

\bibitem{qccp}
\BIBentryALTinterwordspacing
Q.~Du, J.~Xu, Y.~Zhu, H.~Lian, Q.~Xiong, D.~Zheng, Y.~Liu, Z.~Tu, and Z.~Shan,
  ``Qccp: a taskflow programming model for emerging computing scenario,''
  \emph{EPJ Quantum Technology}, vol.~12, no.~1, p.~23, Feb 2025. [Online].
  Available: \url{https://doi.org/10.1140/epjqt/s40507-025-00318-5}
\BIBentrySTDinterwordspacing

\bibitem{openqasm}
\BIBentryALTinterwordspacing
A.~W. Cross, L.~S. Bishop, J.~A. Smolin, and J.~M. Gambetta, ``Open quantum
  assembly language,'' 2017. [Online]. Available:
  \url{https://arxiv.org/abs/1707.03429}
\BIBentrySTDinterwordspacing

\bibitem{gaberle_thesis}
C.~Gaberle, ``Design and implementation of a quantum circuit preparation
  algorithm,'' 2023, unpublished thesis, Goethe University Frankfurt.

\bibitem{tket}
S.~Sivarajah, S.~Dilkes, A.~Cowtan, W.~Simmons, A.~Edgington, and R.~Duncan,
  ``Tket: A retargetable compiler for nisq devices,'' \emph{Quantum Science and
  Technology}, vol.~6, 11 2020.

\bibitem{hpc_with_qpus}
\BIBentryALTinterwordspacing
K.~A. Britt and T.~S. Humble, ``High-performance computing with quantum
  processing units,'' \emph{J. Emerg. Technol. Comput. Syst.}, vol.~13, no.~3,
  Mar. 2017. [Online]. Available: \url{https://doi.org/10.1145/3007651}
\BIBentrySTDinterwordspacing

\bibitem{qiskit}
\BIBentryALTinterwordspacing
A.~Javadi-Abhari, M.~Treinish, K.~Krsulich, C.~J. Wood, J.~Lishman, J.~Gacon,
  S.~Martiel, P.~D. Nation, L.~S. Bishop, A.~W. Cross, B.~R. Johnson, and J.~M.
  Gambetta, ``Quantum computing with qiskit,'' 2024. [Online]. Available:
  \url{https://arxiv.org/abs/2405.08810}
\BIBentrySTDinterwordspacing

\bibitem{demistify_hpc_qc}
\BIBentryALTinterwordspacing
P.~Viviani, ``Demistifying hpc-quantum integration: it's all about
  scheduling,'' in \emph{Proceedings of the 2024 Workshop on High Performance
  and Quantum Computing Integration}, ser. HPQCI '24.\hskip 1em plus 0.5em
  minus 0.4em\relax New York, NY, USA: Association for Computing Machinery,
  2024, p. 1–3. [Online]. Available:
  \url{https://doi.org/10.1145/3659996.3673223}
\BIBentrySTDinterwordspacing

\bibitem{gate_times}
P.~Gerbert and F.~Rue{\ss}, ``The next decade in quantum computing and how to
  play,'' \emph{Boston Consulting Group}, p.~5, 2018.

\bibitem{ibm_devices}
IBM, ``\emph{Compute resources},'' 2025,
  \url{https://quantum.ibm.com/services/resources} [Accessed: Jan 31, 2025].

\bibitem{prace_state_of_the_art}
A.~Tekin, A.~Tuncer~Durak, C.~Piechurski, D.~Kaliszan, F.~Aylin~Sungur,
  F.~Roberts{\'e}n, and P.~Gschwandtner, ``State-of-the-art and trends for
  computing and interconnect network solutions for hpc and ai,'' PRACE, Tech.
  Rep., 2021.

\bibitem{shor_resource_estimation}
\BIBentryALTinterwordspacing
J.~Yamaguchi, M.~Yamazaki, A.~Tabuchi, T.~Honda, T.~Izu, and N.~Kunihiro,
  ``Estimation of shor's circuit for 2048-bit integers based on quantum
  simulator,'' Cryptology {ePrint} Archive, Paper 2023/092, 2023. [Online].
  Available: \url{https://eprint.iacr.org/2023/092}
\BIBentrySTDinterwordspacing

\bibitem{mean-field-ansatz}
\BIBentryALTinterwordspacing
M.~S. Jattana, F.~Jin, H.~De~Raedt, and K.~Michielsen, ``Improved variational
  quantum eigensolver via quasidynamical evolution,'' \emph{Phys. Rev. Appl.},
  vol.~19, p. 024047, Feb 2023. [Online]. Available:
  \url{https://link.aps.org/doi/10.1103/PhysRevApplied.19.024047}
\BIBentrySTDinterwordspacing

\bibitem{nisq_era}
J.~Preskill, ``Quantum computing in the nisq era and beyond,'' \emph{Quantum},
  vol.~2, p.~79, 2018.

\bibitem{vqe_original}
\BIBentryALTinterwordspacing
A.~Peruzzo, J.~McClean, P.~Shadbolt, M.-H. Yung, X.-Q. Zhou, P.~J. Love,
  A.~Aspuru-Guzik, and J.~L. O'Brien, ``A variational eigenvalue solver on a
  photonic quantum processor,'' \emph{Nature Communications}, vol.~5, no.~1, p.
  4213, Jul 2014. [Online]. Available: \url{https://doi.org/10.1038/ncomms5213}
\BIBentrySTDinterwordspacing

\bibitem{vqe_overview}
\BIBentryALTinterwordspacing
J.~Tilly, H.~Chen, S.~Cao, D.~Picozzi, K.~Setia, Y.~Li, E.~Grant, L.~Wossnig,
  I.~Rungger, G.~H. Booth, and J.~Tennyson, ``The variational quantum
  eigensolver: A review of methods and best practices,'' \emph{Physics
  Reports}, vol. 986, pp. 1--128, 2022, the Variational Quantum Eigensolver: a
  review of methods and best practices. [Online]. Available:
  \url{https://www.sciencedirect.com/science/article/pii/S0370157322003118}
\BIBentrySTDinterwordspacing

\bibitem{qaoa_original}
\BIBentryALTinterwordspacing
E.~Farhi, J.~Goldstone, and S.~Gutmann, ``A quantum approximate optimization
  algorithm,'' 2014. [Online]. Available: \url{https://arxiv.org/abs/1411.4028}
\BIBentrySTDinterwordspacing

\bibitem{qaoa_review}
\BIBentryALTinterwordspacing
K.~Blekos, D.~Brand, A.~Ceschini, C.-H. Chou, R.-H. Li, K.~Pandya, and
  A.~Summer, ``A review on quantum approximate optimization algorithm and its
  variants,'' \emph{Physics Reports}, vol. 1068, pp. 1--66, 2024, a review on
  Quantum Approximate Optimization Algorithm and its variants. [Online].
  Available:
  \url{https://www.sciencedirect.com/science/article/pii/S0370157324001078}
\BIBentrySTDinterwordspacing

\bibitem{nlopt}
\BIBentryALTinterwordspacing
S.~G. Johnson, ``The nlopt nonlinear-optimization package,'' 2024, accessed on:
  17. Dec 2024. [Online]. Available: \url{http://github.com/stevengj/nlopt}
\BIBentrySTDinterwordspacing

\bibitem{optimization_direct}
\BIBentryALTinterwordspacing
D.~R. Jones, C.~D. Perttunen, and B.~E. Stuckman, ``Lipschitzian optimization
  without the lipschitz constant,'' \emph{Journal of Optimization Theory and
  Applications}, vol.~79, no.~1, pp. 157--181, Oct 1993. [Online]. Available:
  \url{https://doi.org/10.1007/BF00941892}
\BIBentrySTDinterwordspacing

\bibitem{Garcia2024}
D.~Peral-García, J.~Cruz-Benito, and F.~J. García-Peñalvo, ``Systematic
  literature review: Quantum machine learning and its applications,''
  \emph{Computer Science Review}, vol.~51, p. 100619, 2024.

\bibitem{Matic2022}
A.~Matic, M.~Monnet, J.~Lorenz, B.~Schachtner, and T.~Messerer,
  ``Quantum-classical convolutional neural networks in radiological image
  classification,'' 04 2022.

\bibitem{Dumiak2023}
M.~Dumiak, ``Exascale comes to europe: Germany will host jupiter, europe's
  entry into the realm of exascale supercomputing,'' \emph{IEEE Spectrum},
  vol.~60, no.~1, pp. 50--51, 2023.

\bibitem{HertenSC24}
A.~Herten, S.~Achilles, D.~Alvarez, J.~Badwaik, E.~Behle, M.~Bode, T.~Breuer,
  D.~Caviedes-Voullième, M.~Cherti, A.~Dabah, S.~E. Sayed, W.~Frings,
  A.~Gonzalez-Nicolas, E.~B. Gregory, K.~H. Mood, T.~Hater, J.~Jitsev, C.~M.
  John, J.~H. Meinke, C.~I. Meyer, P.~Mezentsev, J.-O. Mirus, S.~Nassyr,
  C.~Penke, M.~Römmer, U.~Sinha, B.~v.~S. Vieth, O.~Stein, E.~Suarez,
  D.~Willsch, and I.~Zhukov, ``Application-driven exascale: The jupiter
  benchmark suite,'' in \emph{SC24: International Conference for High
  Performance Computing, Networking, Storage and Analysis}, 2024, pp. 1--45.

\bibitem{LeCun1995}
Y.~LeCun, Y.~Bengio \emph{et~al.}, ``Convolutional networks for images, speech,
  and time series,'' \emph{The handbook of brain theory and neural networks},
  vol. 3361, no.~10, p. 1995, 1995.

\bibitem{Krizhevsky2012}
A.~Krizhevsky, I.~Sutskever, and G.~E. Hinton, ``Imagenet classification with
  deep convolutional neural networks,'' \emph{Communications of the ACM},
  vol.~60, pp. 84--90, 2012.

\bibitem{PyTorch2019}
A.~Paszke, S.~Gross, F.~Massa, A.~Lerer, J.~Bradbury, G.~Chanan, T.~Killeen,
  Z.~Lin, N.~Gimelshein, L.~Antiga, A.~Desmaison, A.~Köpf, E.~Yang, Z.~DeVito,
  M.~Raison, A.~Tejani, S.~Chilamkurthy, B.~Steiner, L.~Fang, J.~Bai, and
  S.~Chintala, ``Pytorch: An imperative style, high-performance deep learning
  library,'' 2019.

\bibitem{Deng2012}
L.~Deng, ``The mnist database of handwritten digit images for machine learning
  research,'' \emph{IEEE Signal Processing Magazine}, vol.~29, no.~6, pp.
  141--142, 2012.

\bibitem{adam}
D.~Kingma and J.~Ba, ``Adam: A method for stochastic optimization,''
  \emph{International Conference on Learning Representations}, 12 2014.

\bibitem{em_original}
\BIBentryALTinterwordspacing
Y.~Li and S.~C. Benjamin, ``Efficient variational quantum simulator
  incorporating active error minimization,'' \emph{Phys. Rev. X}, vol.~7, p.
  021050, Jun 2017. [Online]. Available:
  \url{https://link.aps.org/doi/10.1103/PhysRevX.7.021050}
\BIBentrySTDinterwordspacing

\bibitem{em_overview}
Z.~Cai, R.~Babbush, S.~C. Benjamin, S.~Endo, W.~J. Huggins, Y.~Li, J.~R.
  Mcclean, and T.~E. O'Brien, ``Quantum error mitigation,'' \emph{REVIEWS OF
  MODERN PHYSICS}, vol.~95, no.~4, DEC 13 2023.

\bibitem{em_matrix_based_1}
\BIBentryALTinterwordspacing
Y.~Chen, M.~Farahzad, S.~Yoo, and T.-C. Wei, ``Detector tomography on ibm
  quantum computers and mitigation of an imperfect measurement,'' \emph{Phys.
  Rev. A}, vol. 100, p. 052315, Nov 2019. [Online]. Available:
  \url{https://link.aps.org/doi/10.1103/PhysRevA.100.052315}
\BIBentrySTDinterwordspacing

\bibitem{em_matrix_based_2}
M.~S. Jattana, F.~Jin, H.~De~Raedt, and K.~Michielsen, ``General error
  mitigation for quantum circuits,'' \emph{Quantum Information Processing},
  vol.~19, pp. 1--17, 2020.

\bibitem{em_matrix_based_3}
\BIBentryALTinterwordspacing
P.~Döbler, J.~Pflieger, F.~Jin, H.~D. Raedt, K.~Michielsen, T.~Lippert, and
  M.~S. Jattana, ``Scalable general error mitigation for quantum circuits,''
  2024. [Online]. Available: \url{https://arxiv.org/abs/2411.07916}
\BIBentrySTDinterwordspacing

\bibitem{triple_hybrid}
\BIBentryALTinterwordspacing
M.~S. Jattana, ``Quantum annealer accelerates the variational quantum
  eigensolver in a triple-hybrid algorithm,'' \emph{Physica Scripta}, vol.~99,
  no.~9, p. 095117, aug 2024. [Online]. Available:
  \url{https://dx.doi.org/10.1088/1402-4896/ad6aea}
\BIBentrySTDinterwordspacing

\end{thebibliography}

\end{document}